\documentclass[aps,prl,twocolumn,groupedaddress,showpacs,floatfix]
{revtex4}
\usepackage{graphicx}

\begin{document}

\title{Atomic quasi-Bragg diffraction in a magnetic field}
\author{K.\ F.\ E.\ M.\ Domen}
\author{M.\ A.\ H.\ M.\ Jansen}
\author{W.\ van Dijk}
\author{K.\ A.\ H.\ van Leeuwen}
\email[]{K.A.H.v.Leeuwen@tue.nl}
\thanks{This work is financially supported by the Dutch Foundation
for Fundamental Research on Matter (FOM).}
\affiliation{Department of Applied Physics,
Eindhoven University of Technology, P.O. Box 513, 5600 MB
Eindhoven, The Netherlands }


\date{\today}

\begin{abstract}
We report on a new technique to split an atomic beam coherently with
an easily adjustable splitting angle. In our experiment metastable helium atoms in the
$|\{1s2s\}^3{\rm S}_1 \,\, M=1\rangle$ state diffract from a
polarization gradient light field formed by counterpropagating
$\sigma^{+}$ and $\sigma ^{-}$ polarized laser beams in the presence
of a homogeneous magnetic field. In the near-adiabatic regime, energy
conservation allows the resonant exchange between magnetic energy and
kinetic energy. As a consequence, symmetric diffraction of
$|M=0\rangle$ or $|M=-1\rangle$ atoms in a single order is achieved,
where the order can be chosen freely by tuning the magnetic field. We
present experimental results up to 6$^{\rm th}$ order diffraction ($24
\hbar k$ momentum splitting, i.e., $2.21\,{\rm m/s}$ in transverse
velocity) and present a simple theoretical model that stresses the
similarity with conventional Bragg scattering. The resulting device
constitutes a flexible, adjustable, large-angle, three-way coherent
atomic beam splitter with many potential applications in atom optics
and atom interferometry.
\end{abstract}

\maketitle

Coherent beam splitters form an essential element of atom
interferometers.
One way to construct such an
atomic beam splitter is Bragg scattering, where atoms are
diffracted by a standing light wave. The atoms are partially transmitted (zero order
diffraction) and partially diffracted into a single order,
corresponding to specular reflection of the
incoming atoms from the wave fronts of the light. These Bragg beam
splitters have been used successfully to construct Mach-Zehnder
type atom interferometers~\cite{SiuAuLeePRL1995,HagleyPRA2000}.

Diffraction of the atoms can be viewed as resulting from the
subsequent absorption and stimulated emission of photons from the
left and right running components of the standing light wave. This
results in an atomic momentum change of even multiples of the
photon recoil momentum. The restriction to transmission or
reflection can be understood in terms of energy conservation. With
the axial velocity of the atoms much larger than the transverse
velocity, the system can be reduced to a one-dimensional problem.
If the interaction lasts long enough and the switching as
determined by the laser profile and the axial velocity is
sufficiently gradual, the process develops adiabatically. Energy
conservation then requires the square of the transverse momentum
to be conserved. Atomic Bragg scattering has been studied
extensively \cite{MartinPRL1988,SiuAuLeePRA1995,OberthalerPRL1996,
DuerrQSO1996,DuerrPRA1999}. In previous work
\cite{KoolenPRA2002}, we achieved clean, single-order diffraction
up to eighth order. With ultracold caesium atoms, interferometry with
Bragg beam splitting up to twelfth order has been demonstrated
recently~\cite{MuellerPRL2008}.

Here, we report on a novel atomic beam splitter. It combines the
advantages of standard Bragg scattering (adiabatic transfer into a
single, very high diffraction order) with the convenience of
tuning the splitting angle by simply tuning a magnetic field
instead of mechanically adjusting the angle between laser beam and
atomic beam. It also offers the added flexibility of a three-port
beam splitter.

The experimental configuration uses counterpropagating $\sigma^+$
and $\sigma^-$ light beams, to create a quasi-standing light wave
(with strong polarization gradient but no intensity modulation), and
a homogeneous, orthogonal magnetic field. The incoming $J=1$ metastable helium atoms
intersect the light wave perpendicularly.
The waist of the
Gaussian light wave along the axial direction of the atomic beam
and the light intensity are such that the interaction is well outside
the Raman-Nath regime. Although the effective
light shift potentials are much deeper than the recoil energy and we
are thus outside the Bragg regime, the interaction develops smoothly
enough for adiabatic following of instantaneous eigenstates to play a
dominant role~\cite{KellerAPhB1999,MuellerPRA2008}. This forms a clear distinction
with the experiments of Ref.~\cite{PfauPRL1993}, which
are performed in the Raman-Nath regime~\cite{DubetskyPRA2001}.

In our configuration the initial transverse momentum of the atoms is zero. Thus,
the
square of the transverse momentum cannot be conserved in diffraction.
Furthermore, the polarization configuration does not lead to a light
shift grating for a two-level system. Both effects suppress conventional
Bragg scattering.
However, efficient diffraction can still occur through a
mechanism we call quasi-Bragg scattering~\footnote{The name
`quasi-Bragg scattering' is based on the
shared characteristics with regular Bragg scattering: population of a
single diffraction order due to (approximate) energy conservation.
It is used in a more general sense than
in Ref.~\cite{MuellerPRA2008}, where it refers to the transition
regime between resonant Bragg and non-resonant Raman-Nath
diffraction.}. This process is best described by a two-step cycle in a
reference frame with the main
quantization axis along the k-vectors of the light. First,
absorption of a $\sigma^-$ photon from one laser beam followed by
stimulated emission of a $\sigma^+$ photon into the other laser
beam transfers the atom from the $|M=+1\rangle$ state to the $|M=-1\rangle$ state,
while changing the momentum of the atom by $2 \hbar k$. Here, $M$ is
the magnetic quantum number of the atom, and $k=2\pi/ \lambda$ is
the wave number of the light, with $\lambda$ the wavelength of the
light. In the second step, Larmor precession in the transverse
magnetic field rotates the atoms via the $|M=0\rangle$ state back to the
$|M=1\rangle$ state, completing the cycle.

Without magnetic field, the atoms can only return from
$|M=-1\rangle$ to $|M=+1\rangle$ by returning the momentum gained
in the first step to the light field, effectively undoing the
diffraction process. Therefore, this description emphasizes a
fundamental difference from standard Bragg scattering: without the
presence of the transverse magnetic field scattering above first
order is prohibited, even in the Raman-Nath regime.

\begin{figure}[htbp]
 \includegraphics[width=75mm]{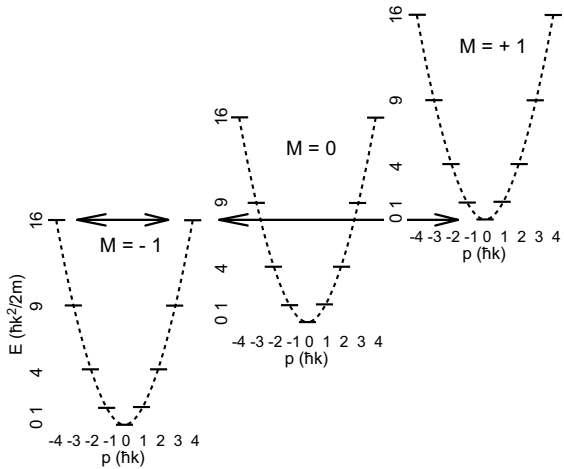}
 \caption{Quadratic kinetic energy potentials $E_{kin}=p^2/2m$
 for the three magnetic substates in a J=1 system.
 In a magnetic field the energy levels of the substates
 are no longer degenerate but shifted by the Zeeman
 interaction ($\Delta E = g_L m \mu_B B$). For
 quasi-Bragg diffraction the magnetic field is tuned
 to balance the increase in transverse kinetic energy. In the figure,
 the resonance between the $|M=1, p=0\hbar k\rangle$ state and the
 $|M=-1, p=\pm 4 \hbar k\rangle$ states is indicated by the arrows.}
 \label{Fig:QBragg_3Potentials}
 \end{figure}

Alternatively, we can also describe the process in a reference frame
with the quantization axis along the magnetic
field. In this frame, the energy conservation criterion is illustrated in Figure
\ref{Fig:QBragg_3Potentials}. The magnetic sublevels are now
non-degenerate.
Outside the Raman-Nath regime, the kinetic energy cannot be neglected.
Therefore we must add to each magnetic substate the kinetic energy
term which is quadratic in the transverse momentum $p$.
For particular values of the magnetic field we can
create degeneracy between Zeeman levels with different transverse
kinetic energy. The light field, in this reference frame a
superposition of $\sigma^+, \sigma^-$ and $\pi$ polarization, then
effectively couples these degenerate eigenstates through
multiphoton Raman transitions.

Efficient transfer to a single diffraction order can now be
achieved, e.g., by starting with atoms in the $|M=1, p=0 \hbar
k\rangle$ substate and tuning the magnetic field such that its
energy equals the kinetic energy of the diffracted states $|M=-1,
p=\pm 2 n \hbar k \rangle$ with $n$ the diffraction order.

In this work, we demonstrate magnetically induced quasi-Bragg
diffraction experimentally. In our setup we produce a monochromatic,
bright, and well-collimated beam of metastable helium atoms.
The beam is collimated by two dimensional laser cooling,
slowed in a Zeeman slower to $247\pm4\,{\rm m/s}$, prefocused by a
magneto-optic lens, and
compressed by a magneto-optic funnel. The design of the beam setup
is described elsewhere \cite{KnopsLPh1999}. After the compression
stage, the beam passes a 25-$\mu$m-diameter aperture $2\,{\rm m}$
downstream. We obtain a flux of 250 atoms per second in the
metastable triplet $^3$S$_1$ state after the aperture with a transverse velocity
spread of $0.05\,{\rm m/s}$. In the experiments described
below, approximately 75$\%$ of the atoms are in the $M=1$ state.

After the collimation aperture the two counter propagating laser
beams ($1.6\,{\rm mm}$ waist radius) with opposite circular
polarization intersect the atomic beam. Typically, the laser
frequency is detuned $1\,{\rm GHz}$ above the resonance with the
$\{1s2s\}^3{\rm S}_1$ to $\{1s2p\}^3{\rm P}_2$ transition at
$1083\,{\rm nm}$ to prevent population of the excited state and
subsequent spontaneous decay. A set of Helmholtz coils enables
nulling of the ambient magnetic field and application of the
desired homogeneous magnetic field at the interaction region
(typically less than one Gauss).

 \begin{figure}[htbp]
 \includegraphics[width=75mm]{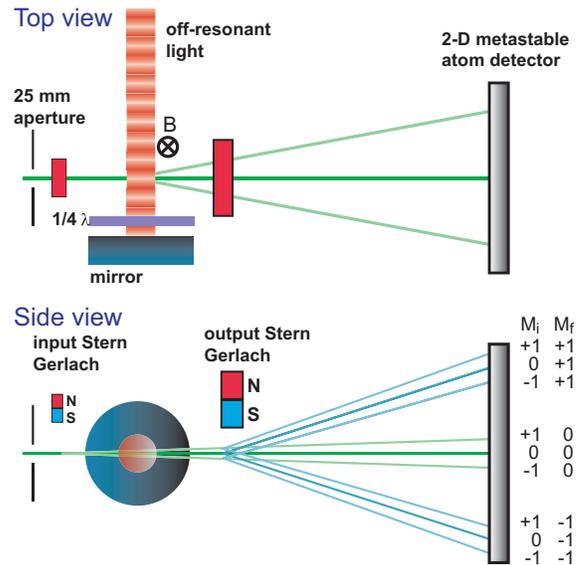}
 \caption{Double Stern Gerlach deflection `tags' each atom
  according to its initial and final magnetic substate. Diffraction
  patterns emerge horizontally, perpendicular to the magnetic
  deflection.}
  \label{Fig:DoubleSternGerlach}
 \end{figure}

Diffraction of the atoms by the light occurs in the horizontal
plane. The horizontal position of the atom
on the 2D position-sensitive detector $2\,{\rm m}$
downstream of the interaction region gives the final momentum state
of the atom after the quasi-Bragg diffraction process.

As quasi-Bragg scattering involves the transition between two
specific magnetic substates, we need a diagnostic tool that can
select and resolve the atom's magnetic state before and after the
interaction. This is achieved by also mapping the magnetic
information onto position information by including Stern-Gerlach
regions with inhomogeneous magnetic fields (see
Fig.~\ref{Fig:DoubleSternGerlach}). These fields are produced by
small permanent magnets. The first magnet is positioned in front
of the interaction region with the light. The gradient of the
magnetic field is vertical. For metastable helium atoms in the
triplet $^3$S$_1$ state this results in 3 distinct vertical
positions on the position sensitive detector, enabling
identification of the atom's initial Zeeman substate. The
separation of the trajectories at the position of the laser beams
is very small compared to the laser waist. Similarly, a second
(stronger) Stern-Gerlach magnet after the interaction region
separates the atoms vertically according to the Zeeman substate
after the interaction. The position of each atom on the detector
thus completely identifies the initial and final $|M,p\rangle$
state, providing a complete characterization of the diffraction
process.

\begin{figure}[htbp]
 \includegraphics[width=75mm]{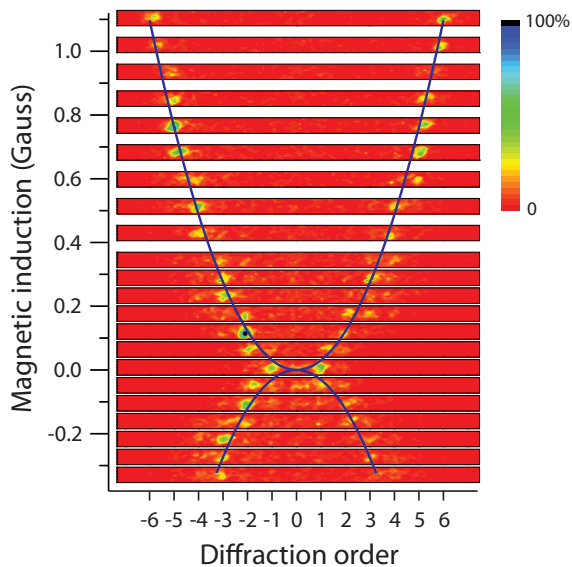}
 \caption{High order quasi-Bragg diffraction as a function of
 applied magnetic field. The spin polarized beam was prepared in
 the $|M=1\rangle$ state with zero transverse momentum. As the
 magnetic field is stepwise increased, the detected diffraction
 patterns for atoms in the $|M=-1\rangle$ output state are displayed.
 The double parabola (solid line) indicates where the magnetic energy
 balances the gain in kinetic energy.}
 \label{Fig:QBS_Result_Diffraction}
 \end{figure}

In the first series of measurements the magnetic field is
increased in small steps while the laser intensity is kept
constant at $51\,{\rm mW}$. The laser frequency is kept at a
detuning $\Delta=1.3\,{\rm GHz}$. We select those atoms from the
data that are initially in the $|M=+1\rangle$ input state and end
up in the $|M=-1\rangle$ output state, as well as the atoms that
end up in the $|M=0\rangle$ state.
Fig.~\ref{Fig:QBS_Result_Diffraction} presents the results of 21
measurements of the $|M=+1\rangle \rightarrow |M=-1\rangle$ atoms.
The vertical position of each of the detector images is centered
around its magnetic field value. At each field strength, the
diffraction pattern shows primarily scattering to one particular
order (plus its mirrored order) that is closest to an energy
resonance as illustrated in Fig.~\ref{Fig:QBragg_3Potentials}.
This is demonstrated by the solid line (double parabola), which
indicates the exact energy resonance. The highest diffraction
order observed (6, corresponding to $24\,\hbar k$ momentum
splitting between the beams) is only limited by the finite
dimensions of the detector. A small fraction of the atoms,
primarily at low magnetic field, is scattered to other diffraction
orders. Approximately 20\% of the atoms undergo spontaneous
emission at the laser power and detuning used.

\begin{figure}[htbp]
 \includegraphics[width=75mm]{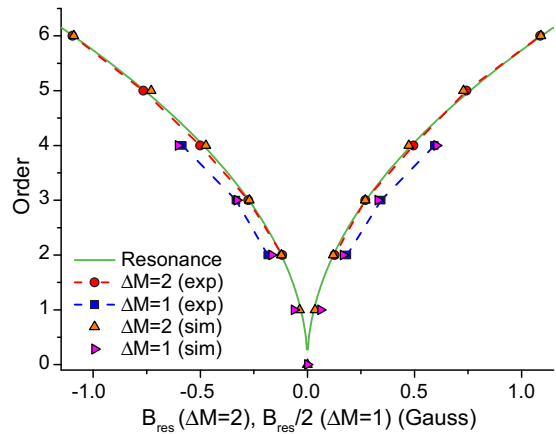}
 \caption{Optimal magnetic field for each observed diffraction order
 for atoms transferred from $|M=1\rangle$ to $|M=-1\rangle$ ($\Delta
 M=2$) as well as for those transferred from $|M=1\rangle$ to
 $|M=0\rangle$ ($\Delta M=1$). The experimental data points are connected
 by dashed lines for clarity. The drawn line indicates where
 the loss in magnetic energy balances the gain in kinetic energy. The
 triangles are the result of numerical simulations. The overall
 calibration factors of the magnetic field and of the laser intensity
 in the presented data are adjusted for best fit between data and
 simulation. The optimum values (fixed for all measurements) are equal
 to independently measured values to well within the estimated $10\%$
 uncertainty interval in the latter.}
 \label{Fig:Dominant-order}
 \end{figure}

From the full set of measurements, the field strengths at which
the transfer is most efficient have been determined for each
diffraction order. The results for both the $|M=+1\rangle
\rightarrow |M=-1\rangle$ and the $|M=+1\rangle \rightarrow
|M=0\rangle$ atoms are plotted in Fig.~\ref{Fig:Dominant-order}.
Both sets of data agree very well with the results of simulations
(indicated by the triangles in the figure). These simulations are
based on direct numerical integration of a time-dependent
Schr\"odinger equation. The quasi-Hamiltonian used includes the
kinetic energy of the atom, the interaction with the Gaussian
shaped light field, as well as a damping term accounting for the
loss by spontaneous emission. The divergence of the incoming
atomic beam is also taken into account in the simulations.

Figure~\ref{Fig:Dominant-order} shows that the $|M=+1\rangle
\rightarrow |M=-1\rangle$ results are very close to the expected
resonance (solid line), whereas the $|M=+1\rangle \rightarrow
|M=0\rangle$ results deviate by 20 to 50\% from the resonant
B-field. These deviations are an indication that quasi-Bragg
scattering is not quite as simple as the basic adiabatic
description given earlier. Detailed analysis shows that in fact
quasi-Bragg scattering is not allowed in the fully adiabatic
limit. Transfer occurs by non-adiabatic Landau-Zener-like transfer
near anticrossings of input and output states that are initially
close in energy, but not degenerate. This mechanism, similar to
the off-resonant Bragg scattering studied in an earlier
paper~\cite{JansenPRA2007} and discussed also by Mueller
\textit{et al.}~\cite{MuellerPRA2008}, will be discussed in a
forthcoming paper.

\begin{figure}[htbp]
  \includegraphics[width=75mm]{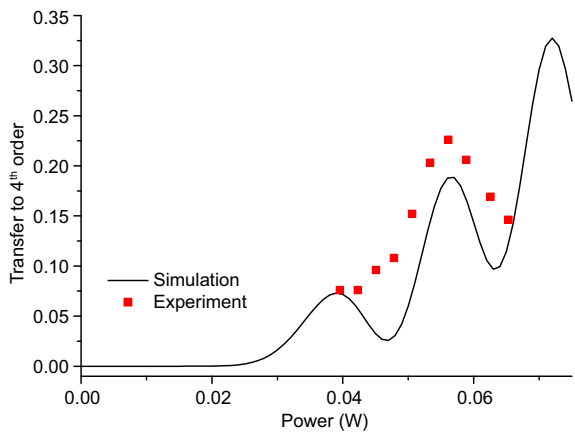}
  \caption{Efficiency of 4$^{\rm th}$ order $|M=1\rangle$ to $|M=0\rangle$
  quasi-Bragg diffraction versus laser power. }
  \label{Fig:Pendellosung}
  \end{figure}

In a second set of measurements the power dependence of the
$|M=1\rangle$ to $|M=0\rangle$ transition was investigated.
Fig.~\ref{Fig:Pendellosung} shows the results. The magnetic field
is kept fixed at resonance for 4$^{\rm{th}}$ order diffraction
($g_L\mu_B B = (8\hbar k)^2/2M_a = 1.10\,{\rm Gauss}$, with $g_L$
the Land\'e factor and $M_a$ the atomic mass) and the detuning at
$\Delta = 1.0$ GHz. The experiments (squares) and simulations,
given by the solid line, are in fair agreement. The results
confirm that the scattered fraction exhibits an oscillatory
behavior similar to the Pendell\"{o}sung oscillations in Bragg
scattering. The small amplitude of the first maximum of the
oscillation in the simulations is a consequence of the
non-adiabatic character of the $|M=1\rangle$ to $|M=0\rangle$
transition.

The limited overall efficiency of the quasi-Bragg process (30$\%$)
is attributable to residual spontaneous emission, reducing the
coherently diffracted population, and to the relatively large
divergence of the atomic beam (0.6 $\hbar k$) as compared to the
expected width of the resonance (0.25 $\hbar k$).

With the data now
available, these issues can be solved by choosing the optimal
intensity and detuning for the desired diffraction order. With an
additional aperture the atomic divergence can be readily reduced
by at least a factor 2 in our setup. Thus, diffraction efficiencies up to 90\%
can be achieved.

To summarize our results: we have demonstrated that clean
high-order atomic diffraction can be achieved in the presence of a
$\sigma^+\sigma^-$ polarized light field and an orthogonal
magnetic field. Key to this quasi-Bragg diffraction is an
efficient and coherent scattering process in which magnetic energy
is converted to kinetic energy. The order of the (symmetric)
diffraction is fully determined by the amplitude of the magnetic field,
obviating the need for mechanical adjustments of the laser-to-atom
angle. The scattering process inherently alters the magnetic substate, which
can be advantageous for high-precision applications (atom
interferometry) where it may be easier to produce a spin-polarized
$|M=\pm1\rangle$ input beam but desirable to have a $|M=0\rangle$
state during the measurements, as this state is to first order
insensitive to stray magnetic fields. The population of the
diffracted order displays an oscillatory behavior
as a function of laser intensity, analogous to the
Pendell\"osung oscillation in conventional Bragg diffraction.


\end{document}